\documentclass[prl,twocolumn,showpacs,superscriptaddress]{revtex4}
\usepackage{epsfig}
\begin{document}

\title{Random Time-Scale Invariant Diffusion and Transport Coefficients}
                                                                                
\author{Y. He}
\affiliation{Department of Physics, Bar Ilan University, Ramat-Gan
52900 Israel}
\author{S. Burov}
\affiliation{Department of Physics, Bar Ilan University, Ramat-Gan
52900 Israel}
\author{R. Metzler}
\affiliation{Physics Department, Technical University of Munich,
D-85747 Garching, Germany}
\author{E. Barkai}
\affiliation{Department of Physics, Bar Ilan University, Ramat-Gan
52900 Israel}

\pacs{02.50.-r,05.40.Fb,87.10.Mn}

\begin{abstract}

Single particle tracking of mRNA molecules and lipid granules in living
cells shows that the time averaged mean squared displacement $\overline{
\delta^2}$ of individual particles remains a random variable while
indicating that the particle motion is subdiffusive. We investigate this
type of ergodicity breaking within the continuous time random walk model
and show that $\overline{\delta^2}$ differs from the corresponding
ensemble average. In particular we derive the distribution
for the fluctuations of the random variable $\overline{\delta^2}$.
Similarly we quantify the response to a constant external field, revealing
a generalization of the Einstein relation. Consequences
for the interpretation of single molecule tracking data
are discussed. 


\end{abstract}
\maketitle

 An ensemble of non interacting Brownian particles spreads according
to Fick's law as a Gaussian packet.
The ensemble averaged mean square displacement (MSD) is $\langle x^2 (t) \rangle
= 2 D_1 t$ where $D_1$ is the diffusion constant. By an Einstein
relation $D_1$ is expressed in terms of statistical
properties of  the microscopic jumps according to
$D_1 = \langle \delta x^2 \rangle / 2 \langle \tau \rangle$ where
$\langle \tau \rangle$ is the average time between jumps and
$\langle \delta x^2 \rangle$ is the variance of the jump lengths. 
Instead one can analyze the time series $x(t)$ 
of the particle trajectory and determine
the time averaged (TA) MSD
\begin{equation}
\overline{\delta^2 } \left(\Delta,t \right) = 
{\int_0 ^{t - \Delta} \left[ x(t' + \Delta) - x (t') \right]^2 {\rm d } t' \over t- \Delta} 
\label{eq01}
\end{equation}
where $\Delta$ is called the lag time.
 For regular Brownian motion and long measurement time
$t \gg  \langle \tau \rangle$ we have 
$\overline{\delta^2 } = 2 D_1 \Delta$, i.e.,
 an ergodic behavior such that
the diffusion coefficient  obtained from an individual
trajectory is identical to the diffusion constant found from
an ensemble of particles under identical physical
conditions.

 From \emph{in vivo\/} single particle tracking the diffusion of lipid
granules in yeast cells \cite{Lenne} and of mRNA molecules in \emph{E.coli\/}
cells \cite{Golding} two findings were made: (i) The TA MSD is subdiffusive,
$\overline{\delta^2 } \sim 2 \overline{D}_{\alpha}
\Delta^{\alpha}$ with $\alpha\approx3/4$.
Usually subdiffusion is defined by the behavior
of an ensemble of particles $\langle x^2 (t) \rangle
=2  D_\alpha t^\alpha/\Gamma(1 + \alpha)$ 
and $0<\alpha < 1$. 
Such anomalous behavior is widespread 
\cite{BouchaudREV,Scher,Metzler,Klages,Condamin},
including charge carrier transport in amorphous semiconductors \cite{Scher},
models of gene regulation \cite{Lomholt},
enzymatic binding in crowded cellular environments \cite{Yutse}
and anomalous dynamics of cell migration \cite{Klages1} 
to name but a few. 
(ii) The second striking observation  \cite{Lenne,Golding} was that 
the TA diffusion coefficient $\overline{D}_\alpha$ is 
a random variable different from the diffusion constant of the
ensemble $D_\alpha$, 
albeit  the measurement time is long (see below) \cite{Platani}. 
Namely using Eq. (\ref{eq01}) to compute a TA 
MSD we get a result which varies from one
single particle trajectory to another \cite{Lenne,Golding} 
(see Fig. \ref{fig5}).
This means that ergodicity is broken such
that time and ensemble averages 
of the diffusion process are non identical. 

In this manuscript we investigate a widely applicable model
for anomalous diffusion: the continuous time random walk (CTRW) 
\cite{BouchaudREV,Scher,Metzler,Klages,Condamin}.
In CTRW subdiffusion is scale invariant and $\langle \tau \rangle \to
\infty$ which naturally leads to ergodicity breaking \cite{WEB,Bel}. 
We show that for the  subdiffusive CTRW the TA MSD (\ref{eq01}) differs
from the ensemble average, even in the limit of long averaging times.  
We obtain the distribution of TA MSDs that completely quantifies the
magnitude of the new fluctuations. Then we treat the biased
random walk 
showing that 
the TA response to an external driving field $F$
also remains random. 
These new
findings   
lead to a new type of
fluctuation-dissipation relation for anomalous kinetics which
depends both on the lag time $\Delta$ and the measurement time $t$.  
Finally we discuss the validity of our theory in experimental
situations and its generality in other models of anomalous diffusion.   

The uncoupled CTRW in one dimension is considered 
\cite{BouchaudREV,Scher,Metzler,Klages}.
The probability density function (PDF) of jump lengths
is $f\left( \delta x \right)$ for which we assume
that its variance  
$\langle\delta x^2\rangle=\int_{-\infty}^{\infty}\delta x^2 f(\delta x)
d(\delta x)$ is finite.
Waiting times between jump events are 
distributed  with a common 
PDF $\psi(\tau)$. 
So the particle waits in its
initial location for a random waiting time, then makes a jump in space,
and then the process is renewed. Our main interest is in the
case where the average sojourn time is infinite 
$\langle \tau \rangle=\infty$ namely
the subdiffusive case with a power law PDF 
$\psi(\tau) \sim A \tau^{ - ( 1 + \alpha)} / |\Gamma( - \alpha)|$ 
and $0<\alpha<1$. 
Physical models which give specific values of $\alpha$ for
different systems and models are given in 
Refs. \cite{BouchaudREV,Scher,Metzler,Klages}.

\begin{figure}
\begin{center}
\epsfxsize=80mm
\epsfbox{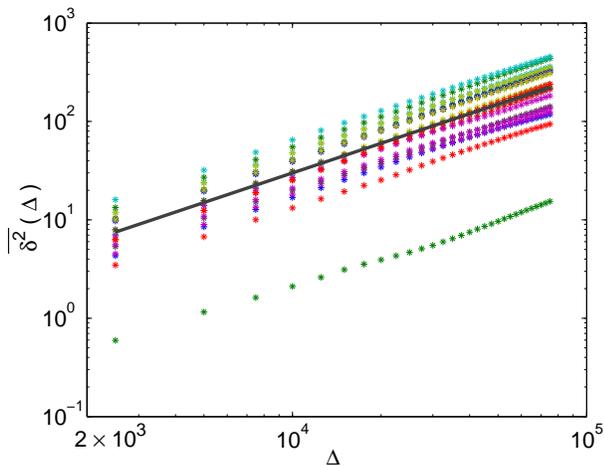}
\end{center}
\caption{
Simulations of the subdiffusive CTRW process with $\alpha=3/4$
and free boundary conditions
show that the TA MSD
is a random variable depending on
individual trajectories.  
The solid curve is the averaged behavior Eq. (\ref{eq04}). 
The measurement time
is  $t=10^8$ and $\psi(\tau) = \alpha \tau^{ - (1 + \alpha)}$ for $\tau>1$.
}
\label{fig1}
\end{figure}

We simulate CTRW trajectories with $\alpha=3/4$, for an unbiased random
walk on a lattice $f(x) = [ \delta(x-1) + \delta(x+1)]/2$
and in Fig. \ref{fig1} show the TA MSD (\ref{eq01}) 
of $10$ individual trajectories with free boundary conditions.   
The most striking feature in the figure is that the
curves are non-identical, and the TA 
MSDs remain a random variable even though a large
number of jump events occur. In contrast if we choose
a waiting time distribution with $\alpha>1$ 
the TAs
will be identical to the ensemble average and non-random 
when  the measurement time is long. 

 To develop a theory for the observed behavior 
 we first consider the average of Eq. (\ref{eq01}) for unbiased
CTRWs, namely for the case when the average jump length is zero.
We consider first free boundary conditions,
a  widely applicable 
case
\cite{BouchaudREV,Scher,Metzler,Klages,Condamin},
not necessarily relevant for
the bounded motion in the cell (see below). 
We have 
$x(t'+\Delta)- x(t')= \sum_{i=1} ^{n_{(t',t'+ \Delta)}} \delta x_i$
 where $\{ \delta x_i\}$ are random  jump lengths, and  
$n_{(t',t'+ \Delta)}$ is the number of jumps in the interval
$(t',t'+\Delta)$. For the unbiased
CTRW the  $\{ \delta x_i \}$s are independent 
random variables 
with zero mean hence 
$\langle [x(t'+\Delta)- x(t')]^2 \rangle=
\langle \delta x^2 \rangle \langle n_{(t',t'+ \Delta)} \rangle$.
The average number of jumps in $(t',t'+\Delta)$
is 
$\langle n_{(t',t'+ \Delta)} \rangle =
 \langle n_{(0,t'+ \Delta)}\rangle-\langle n_{(0,t')}\rangle$.
Using $\langle n_{(0,t')}\rangle \sim t'^{\alpha}/ [A \Gamma(1 + \alpha)]$
\begin{equation}
\langle \overline{\delta^2} \rangle =
{\langle \delta x^2 \rangle \over A \Gamma(1 + \alpha) } {  t^{1 + \alpha} - \Delta^{1 + \alpha} - (t - \Delta)^{1 + \alpha} \over (1 + \alpha) (t - \Delta) }
\label{eq02}
\end{equation}
is obtained 
from Eq. (\ref{eq01}). 
In the limit $\Delta \ll t$ we find
%
%
%
\begin{equation}
\langle \overline{\delta^2}  \rangle \sim {2 D_{\alpha} \over \Gamma(1 + \alpha)}  { \Delta \over t^{1 - \alpha} },
\label{eq04}
\end{equation}
where we used the generalized Einstein relation $D_{\alpha} = \langle
\delta x^2 \rangle / (2 A)$ \cite{BMK}. For $\alpha\ne 1$ Eq.~(\ref{eq04})
is very different from the behavior found for an ensemble,
$\langle  x^2(t) \rangle = 2 D_\alpha t^\alpha/\Gamma(1 + \alpha)$ 
indicating ergodicity breaking. 
Eq. (\ref{eq04}) shows that if we know through measurement the
ensemble averaged anomalous diffusion coefficient
$D_\alpha$
we can determine the single particle trajectory averaged behavior.
 The result Eq. (\ref{eq04}) can be explained by noting that
the longer the process goes on the more likely we are to find long
trapping times of the order of the measurement time (ageing). 
Hence $\langle \overline{\delta^2}\rangle$ decreases when 
measurement time $t$ is increased. 
 Roughly speaking the diffusion constant depends on
time $D(t)\sim {\rm d} \langle x^2 \rangle/ {\rm d} t  \sim t^{\alpha -1}$ and
Eq. (\ref{eq04}) is described by
$\overline{\delta^2} \simeq D(t) \Delta$ 
so a linear dependence on the lag time also seen in the
simulations Fig. \ref{fig1}, is found.

{\em Distribution of $\overline{\delta^2}$}.
  For the CTRW under investigation we still have the
usual scaling of $x^2 \sim N$ with the number of jumps $N$
in $(0,t)$,
however due to the broad distribution of waiting times $N\sim t^\alpha$
and so $x^2 \sim t^\alpha$. Similar scaling
arguments can be
used to analyze the distribution of $\overline{\delta^2}$. 
Assume no jump event occurs between time $t_1$ and $t_2$ and that
$t_2 \gg  t_1 + \Delta$. Then for $t_1 < t' < t_2-\Delta$
we have $[x(t'+\Delta)-x(t')]^2=0$. Since for the scale free dynamics we have 
long sojourn times of the order of the measurement time without any
jump event, $[x(t'+\Delta)-x(t')]^2=0$ for long
renewal periods separated by shorter periods of activity. 
The most important point to realize is that for the process
$[x(t'+\Delta)-x(t')]^2$ the distribution of sojourn times in state
$[x(t'+\Delta)-x(t')]^2=0$ follows the same power law decay as the
original process $x(t)$ with a waiting time
PDF $\psi(\tau) \sim \tau^{ - (1 + \alpha)}$.
This means that when $N$ serves as the operational time
we have normal behavior
\begin{equation}
\overline{\delta^2}\sim CN/t 
\label{eq05}
\end{equation}
where $C$ is a constant independent of $N$
soon
to be determined, and in the denominator we approximate
$t-\Delta\sim t$.
Let $P_N(t)$ be the probability
of making $N$ jumps in the
time interval $(0,t)$ and $\hat{P}_N (u)$ its Laplace
transform.
From the convolution theorem
$\hat{P}_N(u) =
[1 - \hat{\psi}(u)]\exp[N \ln \hat{\psi}(u)]/u$
as well known \cite{BMK}. 
Since we are interested in the long
time behavior only the small $u$ expansion 
$\hat{\psi}(u) \sim 1- A u^\alpha $ is relevant  
and we have 
\begin{equation}
\hat{P}_N(u) \sim A u^{\alpha - 1} \exp( - N A u^\alpha).   
\label{eq06}
\end{equation}
Inverting to the time domain 
\begin{equation}
P_N(t) \sim  {t \over \alpha A^{1/\alpha} N^{1 + 1/\alpha}} l_{\alpha} \left( { t \over A^{1/\alpha} N^{1 /\alpha} } \right)
\label{eq07}
\end{equation}
where
$l_{\alpha}(t)$ is the one sided L\'evy stable PDF, 
whose Laplace pair is 
$\exp( - u^\alpha)$ \cite{Feller,PRE}. 
To find $C$ we note that after
averaging $\langle \overline{\delta^2 } \rangle= 
C \langle N \rangle / t$, using $\langle N \rangle \sim t^\alpha/ A\Gamma(1 + \alpha)$ and  Eq. 
(\ref{eq04}) we have $C= 2 A D_\alpha \Delta$.
By change of variables we
obtain the PDF 
of the dimensionless random variable
$\xi=\overline{\delta^2}/ \langle \overline{\delta^2} \rangle$ 
using Eqs. (\ref{eq05}) and (\ref{eq07})
\begin{equation}
\lim_{t \to \infty} 
\phi_{\alpha} \left( \xi \right) 
={\Gamma^{1/\alpha}\left(1 + \alpha\right)  
\over \alpha \xi^{1 + 1/\alpha}}
l_{\alpha} \left[ {\Gamma^{1/\alpha} \left(1 + \alpha\right) \over \xi^{1 /\alpha} }\right] .
\label{eq08}
\end{equation}
This is one of our main results since it describes the distribution
of a large class of time
average observables, as we soon
show. 
When $\alpha \to 1$ we have an ergodic behavior
$\lim_{\alpha \to 1} \phi_{\alpha} (\xi) = \delta(\xi - 1)$. 
A measure of ergodicity breaking (EB) is the parameter   
\begin{equation}
\mbox{EB}=\lim_{t \to \infty} { \langle \left( \overline{\delta^2} \right)^2\rangle - \langle\overline{\delta^2} \rangle^2 \over \langle \overline{\delta^2} \rangle^2} =
{ 2 \Gamma^2 \left( 1 + \alpha \right) \over \Gamma\left(1 + 2 \alpha\right)} - 1
\label{eq08a}
\end{equation}
which is independent of the lag time $\Delta$ and 
$D_\alpha$.  

\begin{figure}
\begin{center}
\epsfxsize=80mm
\epsfbox{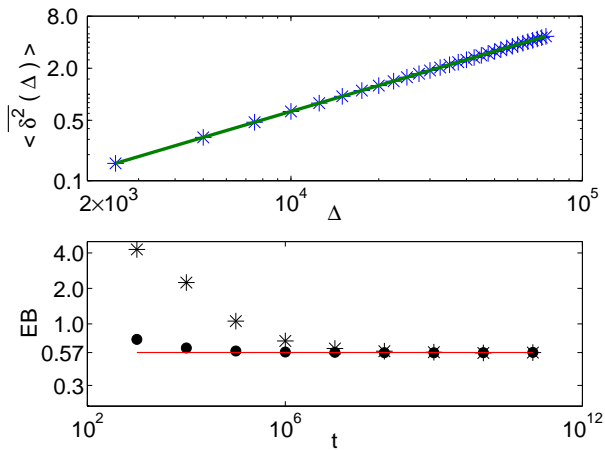}
\end{center}
\caption{
(a) $\langle \overline{\delta^2} \rangle$ versus $\Delta$ for $\alpha=1/2$
and $t=10^8$. Stars 
are simulations and the solid curve
is theory
Eq. (\ref{eq04}) without fitting.
(b) The EB parameter converges slowly to the asymptotic value
$\mbox{EB}=0.5708$ given by Eq. (\ref{eq08a}). 
Here $\Delta=10$ (dots), $\Delta=2500$ (stars)
and $\alpha=1/2$. 
}
\label{fig4}
\end{figure}

\begin{figure}
\begin{center}
\epsfxsize=80mm
\epsfbox{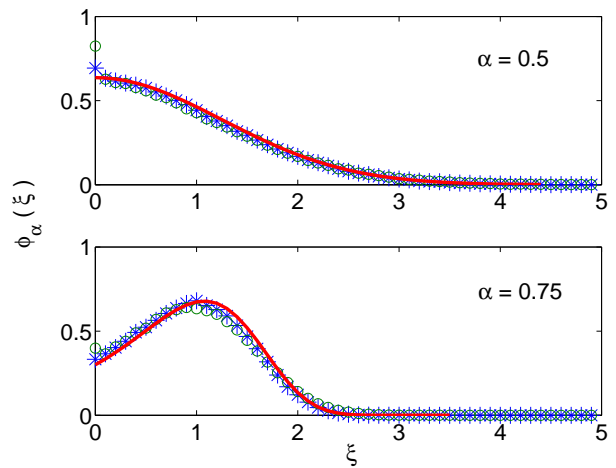}
\end{center}
\caption{
PDF of the scaled random variable 
$\xi=\overline{\delta^2}/\langle \overline{\delta^2} \rangle$ for
 $\alpha=1/2$ and $\alpha=3/4$ with $t=10^8$ and $t=10^7$ respectively.
The full line is  Eq.
(\ref{eq08}).
Stars $(\Delta=2500)$ and circles $(\Delta=4\times 10^4)$
are simulations. 
}
\label{fig3}
\end{figure}

\begin{figure}
\begin{center}
\epsfxsize=80mm
\epsfbox{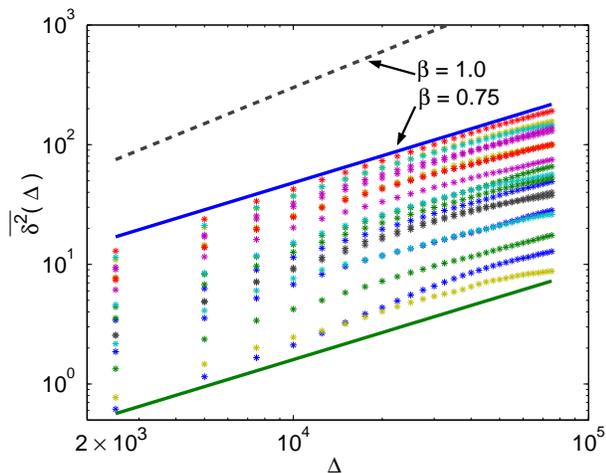}
\end{center}
\caption{Time average $\overline{\delta^2}$ vs. $\Delta$ for unbiased
CTRW on a lattice of size $L=62$, mimicking a particle bounded in a
finite domain, as found in the cell. Compared with the unbounded
case in Fig. \ref{fig1} the diffusion is slower. The trajectory 
 averaged
MSD follows 
$\overline{\delta^2 } \sim \Delta^{\beta}$ 
with $\beta=3/4$ similar to what is 
found in \cite{Golding,Lenne}. The measurement time
was $t=10^8$,  $\alpha=3/4$.}
\label{fig5}
\end{figure}

In Fig. \ref{fig4} we show the behavior of the average
of $\overline{\delta^2}$ and the fluctuations characterized
by the EB parameter, showing excellent agreement between 
asymptotic theory and simulations though the convergence 
of the EB parameter is typically slow.
In Fig. \ref{fig3}
simulations of the PDF of  
$\overline{\delta^2}/\langle \overline{\delta^2} \rangle$ 
for $\alpha=1/2$ and $\alpha=3/4$ are shown. 
We see that for $\alpha=3/4$ we have a peak close to 
$\overline{\delta^2}/\langle \overline{\delta^2} \rangle=1$ 
which indicates that we are closer to the ergodic phase ($\alpha \to 1$)
while for $\alpha=1/2$ the peak is on zero indicating 
stronger non-ergodic behavior as we decrease $\alpha$.

{\em Biased CTRW and Generalized Einstein Relation.}
Now we assume that 
$\langle \delta x \rangle\ne 0$ but constant,
a case which leads to anomalous drift.
 We consider the TA
\begin{equation}
\overline{\delta}\left(\Delta, t \right) =\int_0 ^{t - \Delta}
[ x(t'+ \Delta) - x(t')] {\rm d} t'/( t - \Delta).
\label{eq09}
\end{equation}
First we obtain the average which is done 
with an approach similar to the unbiased case and using
 $\langle x(t'+\Delta) - x(t')\rangle = \langle \delta x \rangle[(t'+\Delta)^\alpha - t'^\alpha]/ [ A \Gamma(1 + \alpha)]$
we find
\begin{equation} 
\langle \overline{\delta} \rangle \sim {\langle \delta x \rangle \over A \Gamma(1 + \alpha)} { \Delta \over t^{1 - \alpha}}.
\label{eq10}
\end{equation}
for $t \gg \Delta$.
Then we can show that the PDF of 
$\xi=\overline{\delta}/\langle \overline{\delta}\rangle$
is given by
Eq. (\ref{eq08}) thus fluctuations of the TA MSD
of the unbiased random walk  and the fluctuations of the 
biased mean response have identical distributions. 

As is
well known according to the generalized Einstein relation 
the transport of an ensemble of particles is related to the
free diffusion of the same particles by 
$\langle x(t) \rangle_F = F \langle x^2 (t) \rangle / 2 k_b T$
where $\langle x^2 (t) \rangle$ is the ensemble average MSD
in the absence of a force field and $\langle x(t) \rangle_F$
is the mean drift when a constant force $F$ is applied to the system
\cite{BouchaudREV,Ein,Grig,Soko}.
This relation can be used to prove that on a microscopic scale
$\langle \delta x \rangle_F = F \langle \delta^2 x \rangle / (2 k_b T)$
which can be obtained from thermal detailed balance conditions \cite{Ein}.
Using this relation and Eqs.  
(\ref{eq04}) and
(\ref{eq10})
\begin{equation}
\langle \overline{\delta} \rangle_F = F \langle \overline{\delta^2} \rangle/
[ 2 k_b T ]
\label{eq11}
\end{equation}
which is valid
under usual linear response assumptions.
This Einstein relation for the TAs  while clearly
related to the Einstein relation for the ensemble average, is valid for
any lag time $\Delta$ and measurement time $t$. In this sense
it differs from the usual Einstein relation. 
As mentioned, the relation between 
transport and diffusion runs deeper, at least within the CTRW
model, since we showed that the fluctuations are identical
as long as the external field does not modify $\psi(\tau)$
(a reasonable assumption for weak fields \cite{Ein})
and described by 
Eq. (\ref{eq08}). 

 {\em Relation with experiments.}
Our results can be tested in single particle
experiments for example for a bead anomalously diffusing
in an actin network which exhibits a CTRW type of dynamics \cite{Weitz}.
However in
the experiments in the cell \cite{Lenne,Golding}
the particle motion
is bounded by the cell walls. Indeed
the  particles may interact with the cell wall
many times whenever $2 D_\alpha t^\alpha/\Gamma(1+ \alpha) > L^2$,
where $L$ is the system length. Finiteness  of the system implies
that at long times the ensemble averaged
MSD will not increase with time but 
rather  saturate.

 We have simulated the effect
of a boundary by considering an unbiased CTRW on a lattice with
system size $L=62$ with lattice spacing equal unity and $\alpha=3/4$.
The simulations shown in Fig. \ref{fig5}
look similar to
experiment and we have $\overline{\delta^2}
\simeq \Delta^\beta$
with $\beta \simeq 3/4$ 
at least within a reasonable time window. 
The exponent $\beta<1$
depends on the system size, on $\alpha$ and on the time
window under investigation. Still our numerical
results show that CTRW theory is compatible with
available experiment. 
A direct test of our theory would be to change the experimental
time $t$ (not only $\Delta$ as done so far) and see if
the TA diffusion slows down with increasing 
$t$. 

 Notice that for free boundary conditions
we get $\overline{\delta^2} \propto \Delta$,
 the existence of the boundaries 
thus causes the diffusion to appear 
slower (i.e. $\beta<1$) which is intuitively
expected.
For free boundary conditions what appears as normal
diffusion 
in a single particle measurement therefore {\it may actually be a
hidden subdiffusive process}. For both free and reflecting boundary
conditions as we increase the measurement time diffusion is slowed
down when the TA procedure is made.
In all such experiments it is thus imperative to
analyze the TA MSD also as function of the measurement time $t$. Additional
clues about the nature of the diffusion are the potential
scatter of the diffusivity as well as the shape of the trajectories.

In the cell, the measurement time $t$
is limited by the life time of the cell.
This is important from the point of view of theory
which usually assumes an infinite measurement time.
Indeed in a finite volume one can expect from
a thermodynamical argument demanding stationarity
that if $\psi(\tau)$ decays  like a power law
it does so only  within a finite time interval and then a cutoff will appear.
However the finite life time of the cell implies that
the usual long time limit essential for ergodicity
may not be reached and ergodicity
breaking is found:   
Ergodicity of diffusion processes is not fulfilled
in a living cell. 

 More generally we expect TA diffusion and transport
coefficients to remain random  
in other models of anomalous diffusion. We have recently
shown \cite{Korabel}
that for intermittent weakly chaotic systems exhibiting
anomalous diffusion, the distribution of scaled 
TA Lyaponov exponents is 
described by Eq.
(\ref{eq08}); similar behavior is
found for super-diffusive L\'evy walks.
For random walks in random environments (e.g.,
random trap and comb models) we expect similar
behavior due to the deep connections
between these models and CTRW theory
(in these models $\langle \tau \rangle \to \infty$ as
in the subdiffusive CTRW however the disorder is quenched not annealed).
Thus one of the  most basic paradigms of transport and diffusion theory,
namely that information obtained from single particle
tracking  is contained already in  the ensemble
measurement is not valid for anomalous diffusion.
This has ramifications for vast classes of processes.

%
This work was supported by the Israel Science Foundation and the Kort
fellowship. We thank Ido Golding and Lene Oddershede for discussions.

\end{document}